\documentstyle[epsfig]{l-aa}
\input epsf

\begin{document}

\thesaurus{12.12.1}

\title{Force equation of the large-scale structure of the Universe}

\author{Mart\'\i n L\'opez-Corredoira}

\offprints{martinlc@iac.es}

\institute{Instituto Astrof\'\i sico de Canarias,
E-38200 La Laguna, Tenerife, Spain}

\date{Accepted xxxx.
      Received xxxx;
      in original form xxxx}
 
\maketitle

\begin{abstract}

This article presents a statistical-mechanical treatment of a relationship (the force equation) 
between the gravitational
potential for two particles and the correlation functions in a relaxed
distribution of particles with different masses. This relationship
is used in the case of galaxies interacting through a Newtonian potential 
in an Universe in expansion, i.e. the large-scale distribution
of galaxies.

By applying this equation and from 
the observed two-point correlation function
for galaxies as a $-1.8$ exponent power law, I derive 
the approximate dependence of a mass-mass correlation
function as a $-2.8$ exponent power law, i.e. I infer that mass is
more correlated than galaxies at short distances, when the distribution
is considered as relaxed.

\end{abstract}

\section{ Introduction}

In a physical system consisting of a collection of particles,
information can be obtained
from the way in which the particles are distributed in space.
For a  Poissonian distribution in equilibrium there is no 
correlation
among the the particles, the position of each one being independent on the position of the
others. This distribution sets the particles randomly in space
with equal probability for all the positions. In contrast, when
the distribution deviates from Poissonian, the correlations among
particles are not null; instead, the particles will be mutually interacting. 

The relationship between the correlations of the distribution and the 
interaction force is called the {\it force equation} and can be obtained
under a certain hypothesis (also called the 
{\it cosmic virial theorem} in the literature relating to large-scale structure;
Peebles 1976). The development of statistical-mechanical theory of liquids has allowed this relationship to be achieved
in the specific case 
where the particles are all equal, point-like 
and in Boltzmann equilibrium, which is quite common for classical particles in
thermodynamic equilibrium (see for 
example Goodstein 1975, March \& Tosi 1976). In general,  statistical-mechanical tools
are applied to obtain the distribution of particles (molecules
on the atomic scale) from an assumed form of the interaction (the Lennard-Jones
interaction among molecules).

Since these statistical applications can be developed in environments other than 
on the atomic scale, my purpose is to derive a few
relationships that will be useful in cosmology.
Groups of galaxies in the large-scale structure of the
Universe are observed in sky surveys, i.e. we have a distribution of particles
in space although there are some differences with respect to the liquid. 
First of all, we know that these
particles are not all identical; their mass and other
characteristics that are not important  dynamically are distinctive among themselves. In view of this,   we could  face 
problems when applying statistical mechanics on
astronomical scales, but the issue of different masses 
is still an important difficulty to resolve first.

Our purpose is to obtain a force equation for a distribution of particles
with different masses that will allow different
magnitudes of interest to be related among themselves. Our result will apply directly to Newtonian gravity for a Universe in expansion.

The necessary condition for solving the problem is to derive information
about: (i) the probabilities of different configurations and
(ii) the Maxwell-Boltzmann equilibrium for classical particles,
which is the case that I will solve. 
This approach was already considered by other authors for the
largest scales in the Universe (e.g. Saslaw \& Hamilton 1984; 
Saslaw 1985) and
I also believe that this condition could be applicable in some gravitational
systems and in some scales
(not all though, see Betancort-Rijo 1988): 
for example, the distribution of galaxies may be a good candidate
to consider relaxed in which we will develop an example of
the expression derived here in Sect. VIII. Nevertheless, there might be other gravitational systems for which this approach may be valid.

\section{ The probability of mass and position for a particle with a known
mass function distribution and Boltzmann equilibrium.}
\begin{description}
\item
{\sc Hypothesis 1:} {\it We have a grand canonical ensemble of
$N$ point-particles, interchangeable with an external reservoir, 
with positions $\vec{r}_1$,...,$\vec{r}_N$ momenta $\vec{p}_1$,...,
$\vec{p}_N$ and
masses $m_1$,...,$m_N$ respectively.
This means that the distribution of positions and momenta
in the particles follows a Maxwell-Boltzmann distribution, i.e.
they are relaxed.}
\end{description}

The probability of a configuration in positions and masses is  
${\cal P}_N(\vec{r}_1,...,\vec{r}_N;
m_1,...,m_N)$. First, we define a general distribution of masses,
and then, once we have the mass of each particle established
a priori, the probability of a particle occupying a position depends on its mass:

\[
{\cal P}_N(\vec{r}_1,...,\vec{r}_N;m_1,...,m_N)
\]\begin{equation}
=
e^{\beta \mu N}
P_N(\vec{r}_1,...,\vec{r}_N/m_1,...m_N)\Phi _N(m_1,...,m_N)
\label{pp} ,\end{equation}
where $\mu $ is the chemical potential, $\beta $ is a constant,
 $\Phi _N$ is the mass distribution function, and
$P_N$ the position distribution function which also depends on
the masses as parameters
(the probability of a configuration of positions is conditioned by
the distribution of masses; the slash stands for ``conditioned'').
We have decomposed the probability ${\cal P}_N$ into the product of
three probabilities: first, the probability of having $N$ particles
in a grand canonical ensemble
(see for example Saslaw 1985, ch. 34); secondly,
the probability of a mass configuration,
$\Phi _N$; and, finally, once we know the mass of each particle,
the probability of the positions configuration. In the following subsections
I obtain $P_N$ and $\Phi _N$.

\subsection{ Probability of positions}

We have assumed that the number of particles
is discrete and that the probability of that they be configured
in a certain way is given by a 
function $P_N(\vec{r}_1,...,\vec{r}_N/m_1,...,m_N)$.

To analyse the form of $P_N$ an assumption is needed on how
position space (or the phase space that includes it) is
populated, and this can be carried out by assuming a state of thermodynamic
equilibrium, which was our first hypothesis.
The Boltzmann distribution is characterized by a configuration probability 
proportional to the negative exponential of the 
Hamiltonian of the system, and the Hamiltonian is the sum
of two terms, one depending on the momenta of
the particles and other
 on the positions of the particles.

The Boltzmann probability distribution function for
a fixed time (see for example Stanley 1971) is

\[
P_N(\vec{r}_1,...,\vec{r}_N;\vec{p}_1,...,\vec{p}_N/m_1,...,m_N)
 \]\[ \propto
 e^{-\beta
{\cal H}(\vec{r}_1,...,\vec{r}_N;\vec{p}_1,...,\vec{p}_N/m_1,...,m_N)}
\]\begin{equation}
=e^{-\beta U(\vec{r}_1,...,\vec{r}_N/m_1,...,m_N)} e^{-\frac{1}{2}\beta
\sum_{i=1}^N \frac{\vec{p}_i^2}{m_i}} ,\label{Boltzmann}\end{equation}
where ${\cal H}$ is the Hamiltonian of the gravitational
system, $U$ the potential energy and $\frac{1}{2}
\sum_{i=1}^N \frac{\vec{p}_i^2}{m_i}$ the kinetic energy.
We integrate over the momentum space $\vec{p}_i$, thus
obtaining the probability for the positions space configuration:

\[
P_N(\vec{r}_1,...,\vec{r}_N/m_1,...,m_N) \propto e^{-\beta U(\vec{r}_1,...,
\vec{r}_N;m_1,...,m_N)} \]\begin{equation}\times 
\int d\vec{p}_i...\int d\vec{p}_N e^{-\frac{1}{2}\beta \sum_{i=1}
^N \frac{p_i^2}{m_i}} . \label{prob}
\end{equation}

The masses $m_i$ and $\beta$ are constant so the integrals 
in (\ref{prob}) are constant and 

\begin{equation}
P(\vec{r}_1,...,\vec{r}_N/m_1,...,m_N)_N \propto e^{-\beta U(\vec{r}_1,...,
\vec{r}_N/m_1,...,m_N)}
\label{p} .\end{equation}

\subsection{ Probability of masses}

Obviously, if we have to take into account the difference of masses
between particles, we must first derive the mass distribution.
The usual tool for doing this is the mass distribution function for a particle, $\phi (m)$, i.e.
the probability of a particle with mass between $m$ and $m+dm$ is
$\phi (m)dm$. Actually, $\phi $ represents an average over space of positions
of the mass distribution. 
We could consider the distribution of mass to vary
at different positions, i.e. the centre of a galaxy cluster with respect to 
another position, an effect which should indeed be taken into account. But
this is wrong. Remember that by means of (\ref{pp}), we first chose the
masses of the particles according to their probabilities, and then
we assigned a position to the particle conditional on having a previously 
established mass.          

So the probability of $N$ particles having masses
$m_1$,...,$m_N$, respectively, is

\begin{equation}
\Phi _N(m_1,...,m_N) \propto \phi (m_1)...\phi(m_N)
\label{phi}.\end{equation}

\subsection{ Normalized probability}

Expressions (\ref{pp}), (\ref{p}) and (\ref{phi}), once normalized, 
lead to

\[
{\cal P}_N(m_1,...,m_N;U(\vec{r}_1,...,\vec{r}_N
/m_1,...,m_N))
\]\begin{equation}=
e^{\beta \mu N}
\phi (m_1)...\phi (m_N)\frac{e^{-\beta U
(\vec{r}_1,...,\vec{r}_N/m_1,...,m_N)}}{Z} ,\label{calp}\end{equation}
and the partition function of this canonical mass-dependent distribution
is given by
 
\[ 
Z=\sum _{N=0}^\infty e^{\beta \mu N}
\int dm_1 ... \int dm_N\int d\vec{r}_1...d\vec{r}_N 
 \]\begin{equation}\times
\phi (m_1)...\phi (m_N)
e^{-\beta U (\vec{r}_1,...,\vec{r}_N/m_1,...,m_N)} 
\label{Z} 
.\end{equation}

\section{ Correlation functions}

The mathematical tools that allow important filtered information to be
extracted
from the distribution of particles are the correlation functions. These will
enable us to ascertain whether the distribution is Poissonian, or how large the
difference is with respect to a Poissonian distribution.

Hereafter, I define certain functions that contain the symbol
$<(\cdots )>$ as meaning an average of the quantity $(\cdots )$. 
There are two different ways of calculating these averages:
(i) by volume, and (ii) statistically-mechanically. I  now describe the two ways:

(i) In this case, the average is calculated according
to the expression

\begin{equation}
<(\cdots )>=\frac{1}{V} \int_V d\vec{r}(\cdots ),
\label{average1}\end{equation} 
where $(\cdots )$ represents the quantity whose average we wish
to obtain. It is clear that we are calculating a volume
average. This is the method that describes 
the distribution macroscopically.
The observational correlations are extracted with this algorithm.

(ii) In this case, the average calculation is conducted
by means of the expression

\[
<(\cdots )>=\sum _{N=0}^\infty \int dm_1...dm_N \int 
d\vec{r}_1...d\vec{r}_N (\cdots) 
 \]\begin{equation} \times
{\cal P}_N(\vec{r}_1,
...,\vec{r}_N;m_1,...,m_N)
,\label{average2}\end{equation}

\noindent where $(\cdots )$ is the same as in (i). This method appeals to the
microscopic properties of the physical system.

The definitions that follow include an expansion with the second averaging method.

\subsection{ Particle-particle two-point correlation function}

The density of the particles is $n(\vec{r})=\sum _{i=1}^N \delta
(\vec{r}-\vec{r}_i)$. 

The two-point correlation function for particles 
without autocorrelation ($i \neq j$ in the following expression) is,
from (\ref{calp}) and (\ref{average2}),

\[
<n(\vec{r})n(\vec{r'})> = \sum _{N=0}^\infty e^{\beta \mu N}
\int dm_1 ... \int dm_N
\int d\vec{r}_1...d\vec{r}_N
 \]\[ \times
 n(\vec{r})n(\vec{r'}) 
\phi (m_1)...\phi (m_N) \frac{e^{-\beta U
(\vec{r}_1,...,\vec{r}_N/m_1,...,m_N)}}{Z}  \]\[
=\sum _{N=0}^\infty e^{\beta \mu N}\int dm_1 ... \int dm_N\int d\vec{r}_1...d\vec{r}_N
 \]\[ \times
\left (\sum_{i,j=1;i\ne j}^N \delta(\vec{r}-\vec{r}_i)
\delta(\vec{r}'-\vec{r}_j) \right ) \]\begin{equation} \times
\phi (m_1)...\phi (m_N) 
\frac{e^{-\beta U(\vec{r}_1,...\vec{r}_N/m_1,...m_N)}}{Z} 
. \end{equation}

In order to simplify the notation, the mass dependence of $U$ will not be indicated in what follows.
A straightforward calculation shows that

\[
<n(\vec{r})n(\vec{r}')>=\sum _{N=0}^\infty e^{\beta \mu N}
\frac{N(N-1)}{Z}\int dm_1 ... \int dm_N
 \]\begin{equation}\times
\int d\vec{r}_3...d\vec{r}_N
\phi (m_1)...\phi (m_N)
e^{-\beta U(\vec{r},\vec{r}',\vec{r}_3,...,\vec{r}_N).}
\label{importante}\end{equation}

\noindent If we assume isotropy, the function only depends on $|\vec{r} - \vec{r}'|$;
I denote this as $<nn>(|\vec{r} - \vec{r}'|)$.

\subsection{ Mass-mass two-point correlation function}

The mass density is $\rho (\vec{r})=\sum _{i=1}^N m_i \delta
(\vec{r}-\vec{r}_i)$. The two-point correlation function for mass
without autocorrelation is

\[
<\rho (\vec{r})\rho (\vec{r}')> = \sum _{N=0}^\infty e^{\beta \mu N}
\int dm_1 ... \int dm_N
\int d\vec{r}_1...d\vec{r}_N 
 \]\[ \times
\rho (\vec{r})\rho (\vec{r'}) 
\phi (m_1)...\phi (m_N) \frac{e^{-\beta U
(\vec{r}_1,...,\vec{r}_N)}}{Z}  \]\[
=\sum _{N=0}^\infty e^{\beta \mu N}\int dm_1 ... \int dm_N\int 
d\vec{r}_1...d\vec{r}_N
 \]\[ \times
\left(\sum_{i,j=1;i\neq j}^N m_i m_j \delta(\vec{r}-\vec{r}_i)
\delta(\vec{r}'-\vec{r}_j) \right)\phi (m_1)...\phi (m_N) 
 \]\[ \times
\frac{e^{-\beta U
(\vec{r}_1,...,\vec{r}_N)}}{Z}
=\sum _{N=0}^\infty e^{\beta \mu N}\frac{N(N-1)}{Z}\int dm_1 ... \int dm_N  \]\begin{equation} \times
\int d\vec{r}_3...d\vec{r}_N
 m_1 m_2 \phi (m_1)...\phi (m_N)
e^{-\beta U(\vec{r},\vec{r}',\vec{r}_3,...,\vec{r}_N)}
.\label{corrrhorho} \end{equation}

\noindent With isotropy, the function depends on $|\vec{r} - \vec{r}'|$.

\subsection{ Mass-particle two-point correlation function}

With the same philosophy in mind, we define other functions as the
two-point correlation function mass-particle:

\[
<\rho (\vec{r})n (\vec{r}')> =
\sum _{N=0}^\infty e^{\beta \mu N} \int dm_1 ... \int dm_N\int d\vec{r}_1...d\vec{r}_N
 \]\[ \times
\left(\sum_{i,j=1;i\neq j}^N m_i \delta(\vec{r}-\vec{r}_i)
\delta(\vec{r}'-\vec{r}_j)\right)
 \]\[ \times 
\phi (m_1)...\phi (m_N) 
\frac{e^{-\beta U
(\vec{r}_1,...,\vec{r}_N)}}{Z}=\sum _{N=0}^\infty e^{\beta \mu N}
\frac{N(N-1)}{Z} 
 \]\[ \times
\int dm_1 ... \int dm_N\int d\vec{r}_3...
d\vec{r}_Nm_1 \phi (m_1)...\phi (m_N)
 \]\begin{equation} \times
e^{-\beta U(\vec{r},\vec{r}',\vec{r}_3,...,\vec{r}_N)}
.\end{equation}

\noindent With isotropy, the function depends on $|\vec{r} - \vec{r}'|$.
 
\subsection{ Mass-mass-particle three-point correlation function}

This function is defined as follows:

\[
<\rho (\vec{r})\rho (\vec{r}')n (\vec{r}'')> 
\]\[ =
\sum _{N=0}^\infty e^{\beta \mu N}
\int dm_1 ... \int dm_N\int d\vec{r}_1...d\vec{r}_N
 \]\[\times
\left(\sum_{^{i,j,k=1}_{i\neq j, j\neq k, k\neq i}}^N m_i m_j \delta(\vec{r}-\vec{r}_i)
\delta(\vec{r}'-\vec{r}_j) \delta(\vec{r}''-\vec{r}_j) \right )
 \]\[ \times
\phi (m_1)...\phi (m_N) \frac{e^{-\beta U
(\vec{r}_1,...,\vec{r}_N)}}{Z} \]
\[
=\sum _{N=0}^\infty e^{\beta \mu N}
\frac{N(N-1)(N-2)}{Z}\int dm_1 ... \int dm_N\int d\vec{r}_4...
 \]\begin{equation} \times
d\vec{r}_N
m_1m_2 \phi (m_1)...\phi (m_N)
e^{-\beta U(\vec{r},\vec{r}',\vec{r}'',\vec{r}_4,...,\vec{r}_N)}
.\label{corrrhorhon}\end{equation}

\noindent With isotropy, this function depends on three variables: 
$|\vec{r} - \vec{r}'|$, $|\vec{r} - \vec{r}''|$ and
$|\vec{r}' - \vec{r}''|$ .

\section{ A force equation for homogeneous, isotropic
 distributions in Boltzmann equilibrium}

In this section, I derive the force equation and then introduce
a Newtonian form for the interaction.

The only trick is to apply the operator $\nabla _{\vec{r}_1}$ over
expression (\ref{importante}) and to develop the expression

\[
\nabla _{\vec{r}_1}<n(\vec{r}_1)n(\vec{r}_2)>=
\sum _{N=0}^\infty e^{\beta \mu N}
\frac{N(N-1)}{Z}\int dm_1 ...
 \]\[ \times
\int dm_N\int d\vec{r}_3...d\vec{r}_N
\phi (m_1)...\phi (m_N)
e^{-\beta U(\vec{r}_1,...,\vec{r}_N)}
 \]\begin{equation}\times
(-\beta) \nabla _{\vec{r}_1} U(\vec{r}_1,...,\vec{r}_N).
\end{equation}
\begin{description}
\item
{\sc Hypothesis 2:} {\it The interaction among particles is
between pairs of particles, and is proportional to their masses,
as a function of their distance (it is a central force).}
\end{description}
By this hypothesis,
the total potential energy is the sum of the potential energies
between pairs of particles ($U(\vec{r}_1,...,\vec{r}_N)=
\sum _{i<j} V_{ij}(\vec {r}_i-\vec {r}_j)$). Hence,

\[
\nabla _{\vec{r}_1}<n(\vec{r}_1)n(\vec{r}_2)>=
\sum _{N=0}^\infty e^{\beta \mu N}
\frac{-\beta N(N-1)}{Z}\int dm_1 ... 
 \]\[ \times 
\int dm_N\int d\vec{r}_3...d\vec{r}_N
\phi (m_1)...\phi (m_N)e^{-\beta U(\vec{r}_1,...,\vec{r}_N)}
 \]\begin{equation}\times
\left[\nabla _{\vec{r}_1}V_{12}(\vec{r}_1-\vec{r}_2) +
\sum _{i=3}^N \nabla _{\vec{r}_1} V_{1i}(\vec{r}_1 - \vec{r}_i)\right].
\end{equation}

\noindent This becomes

\[
\nabla _{\vec{r}_1}<n(\vec{r}_1)n(\vec{r}_2)>=
\sum _{N=0}^\infty e^{\beta \mu N}
\frac{-\beta N(N-1)}{Z}\int dm_1 ... 
 \]\[ \times 
\int dm_N
\nabla _{\vec{r}_1}V_{12}(\vec{r}_1-\vec{r}_2)
\int d\vec{r}_3...d\vec{r}_N \phi (m_1)...\phi (m_N)
 \]\[ \times
e^{-\beta U(\vec{r}_1,...,\vec{r}_N)} 
+\sum _{N=0}^\infty e^{\beta \mu N}\frac{-\beta N(N-1)(N-2)}{Z}
 \]\[ \times
\int dm_1 ... \int dm_N
\int d\vec{r}_3
\nabla _{\vec{r}_1} V_{13}(\vec{r}_1 - \vec{r}_3)
 \]\begin{equation} \times
\int d\vec{r}_4...d\vec{r}_N \phi (m_1)...\phi (m_N) 
e^{-\beta U(\vec{r}_1,...,\vec{r}_N)}.
\end{equation}
\begin{description}
\item
{\sc Hypothesis 3:} {\it We assume homogeneity and isotropy}. 
\end{description}
This means that $<n(\vec{r}_i)n(\vec{r}_j)>$ only
depends on distance $r_{ij}\equiv |\vec{r}_i-\vec{r}_j|$ and
$V(\vec{r})=V(r)$. The operator $\nabla _{\vec{r}_1}$ can be expanded
and thus (I note $<n_1n_2>$ instead of $<nn>$ to avoid confusion;
the same with $<\rho \rho >$ and $<\rho \rho n>$ in the following equations)

\[
r_{12}\frac{\partial <n_1n_2>(r_{12})}{\partial r_{12}}=
\sum _{N=0}^\infty e^{\beta \mu N}\frac{-\beta N(N-1)}{Z}
 \]\[ \times
\int dm_1 ... \int dm_N
[r_{12}\frac{\partial V_{12}(r_{12})}{\partial r_{12}}]
\int d\vec{r}_3...d\vec{r}_N 
 \]\[ \times
\phi (m_1)...\phi (m_N) 
e^{-\beta U(\vec{r}_1,...,\vec{r}_N)} 
+\sum _{N=0}^\infty e^{\beta \mu N}
 \]\[ \times
\frac{-\beta N(N-1)(N-2)}{Z}\int d\vec{r}_3 \int dm_1 ... \int dm_N
 \]\[ \times 
\left[r_{12} \cos (\vec{r}_{12},\vec{r}_{13}) 
\frac{\partial V_{13}(r_{13})}{\partial r_{13}}\right]
 \]\begin{equation} \times
\int d\vec{r}_4...d\vec{r}_N \phi (m_1)...\phi (m_N) 
e^{-\beta U(\vec{r}_1,...,\vec{r}_N)}.
\end{equation}

I rename $r=r_{12}$, $s=r_{13}$, $\theta \equiv$ angle between
$\vec{r}_{12}$ and $\vec{r}_{13}$ and assume that $V_{ij}(r)=m_im_jv(r)$
(valid in the light of Hypothesis 2) and substitute some expressions
using (\ref{corrrhorho}) and (\ref{corrrhorhon}), taking into account
the isotropy assumed in Hypothesis 3:

\[
\frac{-1}{\beta }\frac{\partial <n_1n_2>(r)}{\partial r}=
\frac{\partial v(r)}{\partial r}<\rho \rho >(r) 
\]\[ +
 2\pi \int _0^{\pi } d\theta
\sin \theta \cos \theta \int_0^\infty ds \ s^2 
 \]\begin{equation} \times
\frac{\partial v(s)}{\partial s}
 <\rho _1\rho _2n_3>(r,s,\sqrt{r^2+s^2-2rs\cos \theta })
.\label{forceeq}\end{equation}
This is the force equation. It gives us a relationship between
different correlation functions in a distribution that follows
our hypothesis and the force that is represented by means of $v$.

\section{ Newtonian interaction and expansion}

\subsection{Newtonian interaction}

Before proceeding, I will comment on some aspects regarding the 
Newtonian case in particular. Certain problems are associated with the application
of principles such as thermodynamics and statistical mechanics in
groups of particles with $r^{-2}$-type force (Taff 1985): some
divergences are found, and there is non-saturation of gravitational
forces (Levy-Leblond 1969). Some authors take the view that there can be no
rigorous basis for applying statistical mechanics in such a system 
(Fisher \& Ruelle 1966). This 
result is not a consequence of the $r^{-2}$-type force but rather
of its unshielded character (see Dyson \& Lenard 1967 for a discussion of
the electrostatic case).

In any event, one avoids the non-self-consistency of the problem by
truncating the integral limits at a finite radius, or by cancelling the
correlation functions that fall under a given lower unit, assuming points with 
negligible volume as is the case in real physical problems. In my opinion,  $N$-body
systems exist in nature under Newtonian gravitational forces, and
to invoke a distribution of particles in such systems is not necessarily
inconsistent. The possible  infinities that appear in some expressions
are only mathematical problems which are not present in nature
and can be solved once we give our data
the conditions that will accommodate our physical reality to a mathematical model
(I mean to avoid the infinite proximity of particles by means of a cut-off, etc.). We can look at it from another point of view: a statistical
thermodynamical system in equilibrium cannot achieve singular states 
with infinities, except as a set of zero measurements, because the probability of
getting a singular state, with a very small distance between some particles,
would take an extremely long time, as if approaching a singularity. 
 
Expression (\ref{forceeq}) has been obtained regardless of the force type,
so it possesses a general validity. Now, when we introduce a Newtonian gravitational
force the expression continues to be valid. If an infinity appears in the 
next expression it is only a question of truncating the integrals or selecting
the best correlation function that does not produce divergences.
When we take the cut-off, we neglect the probabilities
near the singularity, a set with dimensions greater than zero but small enough.
The introduction of cut-offs will make the results
dependent upon the details of the regularization, so the selection of
the cut-offs must have a physical basis.
\begin{description}
\item
{\sc Hypothesis 4:} {\it The interaction between pairs of particles is
the Newtonian gravity force}.
\end{description}
This hypothesis obviously includes Hypothesis 2.
With a Newtonian potential: $V_{ij}(\vec{r}_k-\vec{r}_l)=-\frac{Gm_im_j}
{r_{kl}}$. Then, 

\[
\frac{\partial v(r)}{\partial r}=\frac{G}{r^2}.\]

We set $C \equiv (\beta G)^{-1}$. This leads to

\[
-C\frac{\partial <n_1n_2>(r)}{\partial r}=
\frac{1}{r^2}<\rho _1 \rho _2>(r) 
\]\[ +
 2\pi \int _0^{\pi } d\theta
\sin \theta \cos \theta \int_0^\infty ds 
 \]\begin{equation} \times
<\rho _1 \rho _2 n_3>(r,s,
\sqrt{r^2+s^2-2rs\cos \theta })
,\label{testeq}\end{equation}
which gives us a relationship among certain correlations of the distribution
and $C$.

The proof that the system is valid for achieving thermodynamic equilibrium
can be found in Lieb \& Lebowitz (1973), where a general Coulombian system is
considered. Nevertheless, this does not imply that all Newtonian gravitational
systems are in equilibrium. 

\subsection{Effects of the expansion of the Universe}

When we consider the galaxy distribution 
in the large-scale structure of the Universe, we must
bear in mind the expansion, so:
\begin{description}
\item
{\sc Hypothesis 5:} {\it The system of particles is distributed over a space in expansion.}
\end{description}
We must use comoving coordinates in order to maintain
the zero peculiar velocity as the most probable one
(because the above formulation gives the probability as proportional
to $e^{-p}$, where $p$ is the momentum).
We consider the proper motions, not the background: 
the kinetic energy derived from
peculiar velocities and the comoving potential energy.

It is derived in Saslaw \& Fang (1996) that consideration of the expansion
is equivalent to taking into account the gravitational effects
of the local fluctuating part of the density field
when we use comoving coordinates, i.e.
we should substract the mean density from the density field.
Since the mean density does not depend on $r$ or
$\theta $ and the integral $ \int _0^{\pi } d\theta
\sin \theta \cos \theta=0$, equation (\ref{testeq}) 
is modified by simply replacing $<\rho _1 \rho _2>(r)$ 
by $<\rho _1 \rho _2>(r)- <\rho >^2$:

\[
-C\frac{\partial <n_1n_2>(r)}{\partial r}=
\frac{1}{r^2}(<\rho _1 \rho _2>(r)-<\rho >^2) 
\]\[ +
 2\pi \int _0^{\pi } d\theta
\sin \theta \cos \theta \int_0^\infty ds 
 \]\begin{equation} \times
<\rho _1 \rho _2 n_3>(r,s,
\sqrt{r^2+s^2-2rs\cos \theta })
.\label{testeq_exp}\end{equation}

\section{ The meaning of $\beta$}

We know the meaning of $\beta$ in statistical physics; it is equal to $\frac{1}{K_{\rm B}T}$, where $K_{\rm B}$ is
Boltzmann's constant, and $T$ is the absolute temperature.
In the large-scale (or any other) structure we can establish a
similar meaning by developing a kinetic theory, 
in which the particles are the galaxies themselves (or other particles).

If we integrate expression
(\ref{Boltzmann}) over position space, we get

\[
{\cal P}(\vec{p}_1,...,\vec{p}_N;m_1,...m_N) 
\]\begin{equation} \propto
\phi (m_1)...\phi (m_N) e^{-\frac{1}{2}\beta
\sum_{i=1}^N \frac{p_i^2}{m_i}}.
\end{equation}

The mean value of the peculiar velocities for all clustering galaxies
is

\[
\overline{v}=
\frac{\int dm \phi(m) \int d\vec{p}\frac{p}{m}e^{-\frac{1}{2}\beta \frac{p^2}{m}}}
{\int dm \phi(m) \int d\vec{p}e^{-\frac{1}{2}\beta \frac{p^2}{m}}}\]
\begin{equation}
=\frac{ \int dm \frac{\phi (m)}{m} \frac{8\pi m^2}{\beta^2}}{\int dm \phi(m)                 \left(\frac{2\pi m}{\beta}
\right)^{3/2}}=\left(\frac{8}{\pi \beta}\right)^{1/2}
\frac{\int dm \phi(m) m}{\int dm \phi(m) m^{3/2}}
.\end{equation}

We can also express this as

\begin{equation}
\overline{v}=\left(\frac{8GC}{\pi \mu }\right)^{1/2}
,\label{vmean}\end{equation}
where $\mu=\left(\frac{\int dm \phi(m) m}{\int dm \phi(m) m^{3/2}}
\right)^{-2}$ and $C$ represent the same constant as in the force equation
for the Newtonian case.
Hence, note that the parameter $C$, which we could obtain in the equality
(\ref{testeq_exp}), will give information about the velocity field of
the particles, or, vice-versa, we could obtain $C$ for (\ref{testeq_exp})
from the mean velocity. 

The parameter $\beta$ is even more directly related to the kinetic energy. An analogous derivation leads to

\begin{equation}
E_{\rm kin}=\frac{3N}{2\beta}=\frac{3}{2}N\ G\ C
,\end{equation}
a fundamental result of statistical mechanics.
\section{ How to obtain the correlations from the distribution}

To this end, we have to use 
eq. (\ref{average1}) as defined previously. However, 
when we have a discrete number of points instead
of a continuum distribution, perhaps it might be better to use other
equivalent expressions.

When homogeneity is given, one method discussed by Rivolo (1986) is to use the estimator

\begin{equation}
<n_1n_2>(r)=\frac{<n>}{N}\sum_{i=1}^N \frac{N_i(r)}{V_i(r)},
\end{equation}
where $N_i (r)$ is the number of particles lying in a shell of thickness
$\delta r$ from the $i$th particle, $V_i (r)$ is the volume of the shell
lying within the sample volume and $<n>$ is the average density in 
a macroscopically homogeneous system ($<n>$ is independent of position).

The evaluation of $<\rho _1 \rho _2>$ and $<\rho n>$ would be:

\begin{equation}
<\rho _1 \rho _2>(r)=\frac{<\rho >}{M}\sum_{i=1}^N \frac{m_iM_i(r)}{V_i(r)}
\end{equation}
and
\begin{equation}
<\rho _1n_2>(r)=\frac{<n>}{M}\sum_{i=1}^N \frac{m_iN_i(r)}{V_i(r)},
\end{equation}
where $M=\sum _{i=1}^N m_i$ and $M_i (r)$ is the mass 
lying in a shell of thickness $\delta r$ from the $i$th particle.
Knowledge of $m_i$ is sometimes problematic and its solution differs
in each case. In the case of galaxies or stars information is required 
about the mass-luminosity relationship as well as data 
on magnitudes and distances. This could also be provided
with a knowledge of the mass distribution function according to
different zones, ( $\phi (m)$ would be the average of the different
mass distribution functions in the entire space) and by randomly assigning 
a mass $m_i$ to each particle following these distribution functions that
would give us not the real mass distribution but an equivalent one.
In any case, the problem of assigning mass is different in each case and
generally requires tailor-made solutions for each one.

The three-point correlation function is also obtainable by counting
groups of three particles with different distances between them. 
Usually, in the isotropic
case, this is approximated as a function of different two-point
correlation functions. 
For example, in liquid theory the so-called superposition
approximation is commonplace (see for example March \& Tosi 1976 and first
formulation of it in Kirkwood 1935), and applied
to the mass-mass-particle case would take the form

\[
<\rho _1 \rho _2 n_3>(r,s,t)
\]\begin{equation}=
\frac{<\rho _1\rho _2>(r)<\rho _1n_2>(s)<\rho _2 n_3>(t)}
{<\rho>^2<n>}.
\label{superpos}\end{equation}

From the large-scale distribution of galaxies in the Universe, we
can extract statistical information (as in Saunders et al. 1991). When we observe
the projected distribution onto a 2-dimensional surface, i.e. we do not know
the distance of the objects, we can obtain the correlation on the 2-dimensional
surface (angular correlation) and relate it to the 3-dimensional distribution
correlations by means of Limber's equation (Peebles 1980).
Maddox et al. (1990) obtained the two-point angular correlation
function on large scales, so we can derive $<n_1n_2>$. 
To assign the masses in order to achieve
$<\rho _1\rho _2>$ and $<\rho _1n_2>$ is again troublesome and depends on the 
data available on the masses of the galaxies.

For the mass of galaxies in the large-scale structure, future data will become available
with the help of the DENIS project (Mamon 1995), a near-infrared sky survey.
The near-infrared is thought to be a better tracer of the stellar mass in the
galaxies, and if the stellar mass content follows the total mass content (including dark 
matter), near-infrared surveys should be the best way of obtaining the distribution
of matter in the Universe.

Also, on other astronomical scales we find point-particle distributions
and we may obtain their correlations. Borgani et al. (1991) obtained the
correlation functions for scales between 3 and 350 kpc, and
this could be done for smaller scales as well. Another problem is posed by considering
our hypotheses as valid, especially Hypothesis 1.

An application to any other distribution is also possible.

\section{Application on the Large-scale distribution of galaxies}

The actual application of these tools is quite difficult since real data are not easily available.
To illustrate the way to proceed with this method, we apply the above theoretical
results to a practical case with observational data, 
in the $N$-body system of the large scale distribution
of galaxies in the Universe
(see for example Peebles 1980; Borgani 1995) which is a homogeneous and isotropic distribution. 
A recent
model, using equilibrium statistical mechanics as well as other considerations,
was also developed in P\'erez-Mercader et al. (1996).

We will assume the validity of the assumptions made in this paper and
adopt an additional assumption here for the particular case of
the use of (\ref{superpos}), the superposition approximation, and
biasing of the $\rho $ fluctuation as proportional to the $n$ fluctuation
where the constant of proportionality depends on the scale:

\begin{equation}
\frac{\delta n}{n}=b(r)\frac{\delta \rho}{\rho},
\end{equation}
where $b$ is the biasing parameter, dependent a priori on $r$,
the distance between both particles.
Astronomers usually make a stronger assumption by taking $b$ to be constant, 
but I am not going to be so restrictive.
I adopt a unit system where $<\rho >=<n>$ and set $<n_1n_2>(r)\equiv 
<n>^2(1+\xi (r))$.
As a consequence,

\begin{equation}
<\rho _1n_2>=<n>^2(1+b^{-1}\xi (r))
\end{equation}
and

\begin{equation}
<\rho _1\rho _2>=<n>^2(1+b^{-2}\xi (r))
.\end{equation}
Thus, (\ref{testeq_exp}) leads to:

\[
-C\xi '(r)=\frac{b^{-2}(r)\xi (r)}{r^2}
\]\[+
2\pi <n>
\int _0^{\infty} ds[1+b^{-1}(s)\xi(s)+b^{-2}(r)\xi (r)
 \]\[
+b^{-2}(r)b^{-1}(s)\xi (r) \xi(s)] 
\int _0^{\pi }d\theta
\sin \theta \cos \theta 
\]\begin{equation}\times
 b^{-1}(\sqrt{r^2+s^2-2rs\cos\theta })
\xi (\sqrt{r^2+s^2-2rs\cos\theta }).
\label{ex_eq}\end{equation}
This expression resembles the one given by Peebles (1976) for $b=1$
, the cosmic virial theorem,
but with further generality because expression (\ref{ex_eq}) takes into account
that $<~n_1n_2>(r)\ne <\rho _1\rho _2>(r)$.

An observational $\xi $ was achieved by Groth \& Peebles (1977):

\begin{equation}
\xi (r)=\left(\frac{r}{r_0}\right )^{-\gamma} \ {\rm for} \ 
0.3h^{-1}\ {\rm Mpc} < r<10h^{-1}\ {\rm Mpc}
\end{equation}
where $r_0=4.1h^{-1}$ Mpc and  $\gamma =1.77$. The parameter $\xi $ is negligible for $r>10h^{-1}$ Mpc, so it is taken as zero. Also a cut-off is taken
for $r<0.3h^{-1}$ Mpc because separated galaxies cannot be at 
distances less than 
this cut-off (if two galaxies have a distance between themselves less than this then
they are considered as only one galaxy),
and this ensures convergence of the integral in (\ref{ex_eq}).

In order to estimate the behaviour of $b(r)$, we introduce it as
a power-law dependence, such that 

\begin{equation}
b(r)=\left(\frac{r}{r_*}\right)^\kappa
\end{equation}
and fit the best values of $\kappa$ and $r_*$ to solve 
(\ref{ex_eq}). The constants are derived from
observational data:
$<n>=0.02\ {\rm Mpc}^{-3}$ (Allen 1973),
$h=0.60$ (the actual value of the Hubble constant in units of 100 km s$^{-1}$ Mpc$^{-1}$)
and $\overline{v}\sim 100$ km s$^{-1}$ (Allen 1973). 

If we now wish to obtain $C$, according to (\ref{vmean}),
we must convert $G$ from our units 
(time unit: second; length unit: $1$ $h^{-1}$ Mpc;
mass unit: the average mass of a galaxy, $\overline{m}_{\rm gal}$ in K`kg)
to the MKS system, multiplying by
a factor of $(3.08 \times 10^{22})^{-3}
\overline{m}_{\rm gal}$. Thus, denoting by $G_{\rm MKS}$ the
value of $G$ in the MKS system, we have

\begin{equation}
\overline{v}=\left(\frac{8G_{\rm MKS}(3.08 \times 10^{22})^{-3}
\overline{m}_{\rm gal}C}{\pi}\right)^{1/2}\
h^{-1}\ {\rm Mpc\ s}^{-1}
,\end{equation}
so with $G_{\rm MKS}=6.672\times 10^{11}$, $ M_{\odot}=1.99\times 
10^{30}\ {\rm kg}$ and $1 \ {\rm Mpc}=3.08\times 10^{19}\ {\rm km}$

\begin{equation}
C\approx \frac{3.45 \times 10^{13}\ M_{\odot}}{\overline{m}}.
\end{equation}
This, together with an estimate of 
$\overline{m} \approx 8\times 10^{10} \  M_{\odot}$ (Allen 1973), gives us

\begin{equation}
C\approx 4.3 \times 10^2.
\end{equation}

The introduction of all these data into (\ref{ex_eq}) and some calculations
lead to the following expression for the allowed $r$ values: 

{\scriptsize
\[
9.25\times 10^3r^{-2.77}-12.1r_*^{2\kappa}r^{-3.77-2\kappa}=
\frac{0.291r_*^\kappa}{r}
\]\[\times
\int _0^{r+10}\frac{ds}{s}
[
1+12.1r_*^\kappa
\left \{ \begin{array}{ll} 
-1 &\mbox{$ s< 0.3$} \\
s^{-1.77-\kappa} &\mbox{$ 0.3< s< 10$}\\
0 & \mbox{$ s>10$}\\
\end{array} \right \}
\]\[ +
12.1r_*^{2\kappa}r^{-1.77-2\kappa}+83.1r_*^{3\kappa}
\left \{ \begin{array}{ll} 
-1 &\mbox{$ s< 0.3$} \\
s^{-1.77-\kappa} &\mbox{$ 0.3< s< 10$}\\
0 & \mbox{$ s>10$}\\
\end{array} \right \}
]
\]\[ \times
[
\frac{1}{2.23-\kappa}\left(
\left \{ \begin{array}{ll}
0.3^{2.23-\kappa} &\mbox{$ |r-s|< 0.3$} \\
|r-s|^{2.23-\kappa} &\mbox{$ |r-s|>0.3$} \\
\end{array} \right \}
- (r+s)^{2.23-\gamma}
\right) 
\]\begin{equation}-
\frac{r^2+s^2}{0.23-\kappa}\left(
\left \{ \begin{array}{ll}
0.3^{0.23-\kappa} &\mbox{$ |r-s|< 0.3$} \\
|r-s|^{0.23-\kappa} &\mbox{$ |r-s|>0.3$} \\
\end{array} \right \}
-(r+s)^{0.23-\gamma}
\right)
]
\label{ex_eq2}.
\end{equation}

}

\begin{figure}
\begin{center}
\mbox{\epsfig{file=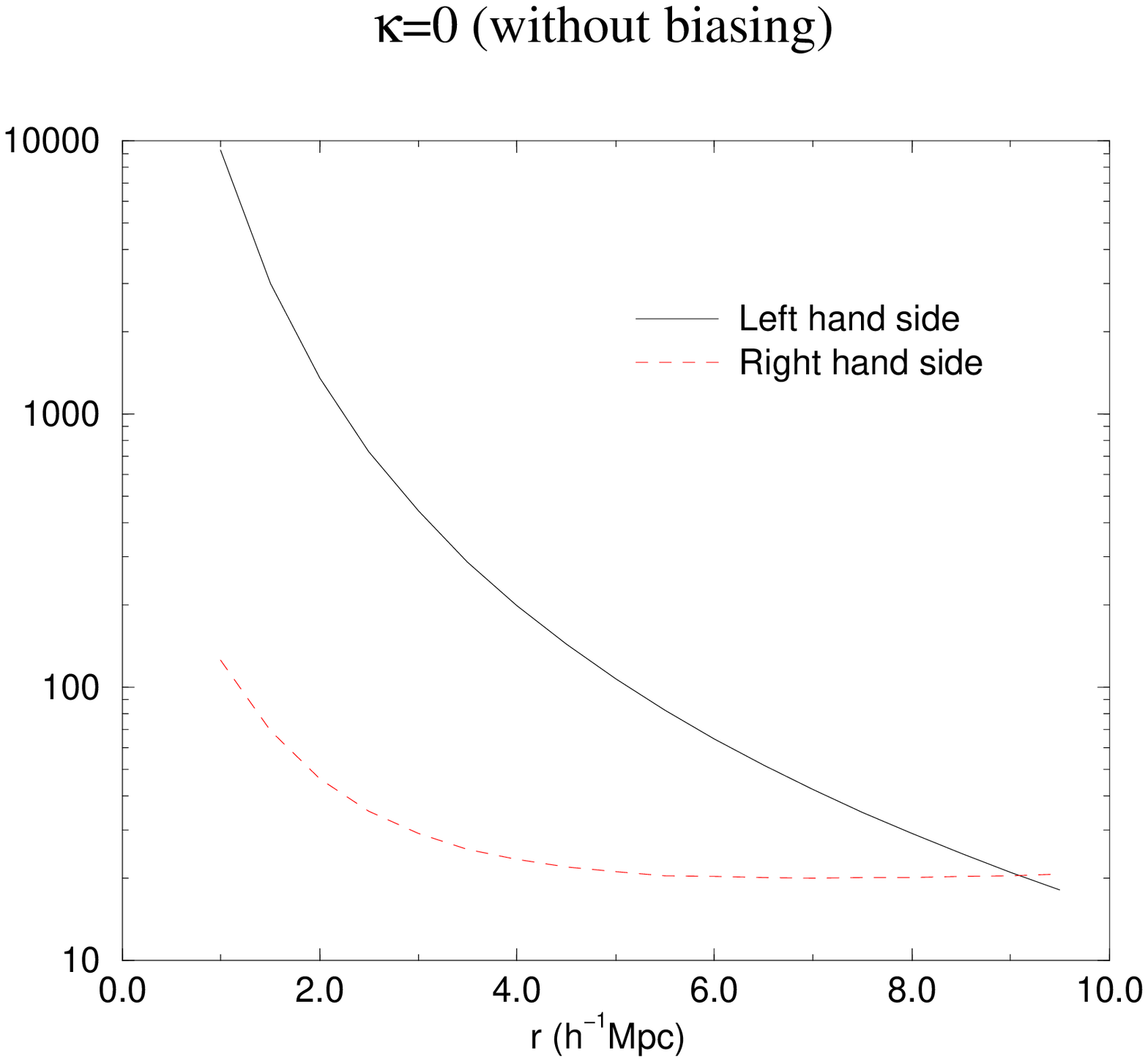,height=7cm}}
\end{center}
\caption{Comparison between the left- and right-hand side of 
expression (\protect{\ref{ex_eq2}}) for the non-biasing case.}
\end{figure}

If non-biasing, i.e. $b=1$, were assumed for all scales, we would obtain
the results plotted in Fig. 1, so biasing is neccesary.

\begin{figure}
\begin{center}
\mbox{\epsfig{file=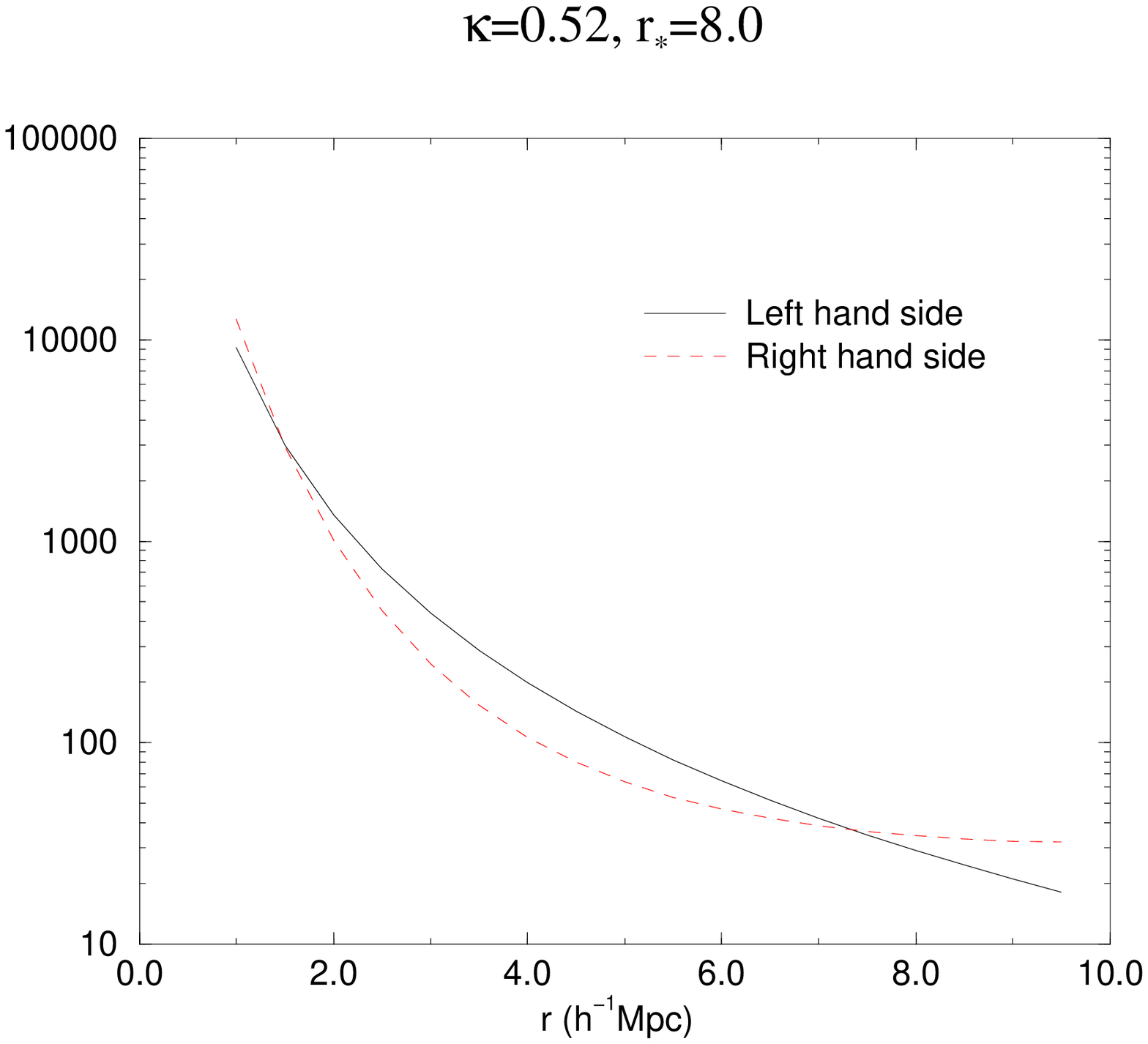,height=7cm}}
\end{center}
\caption{Comparison between the left- and right-hand side of 
expression (\protect{\ref{ex_eq2}}) for best fit of $\kappa$ and $r_*$.}
\end{figure}

The best fit, calculated numerically, is shown in Fig. 2, where left- and right-hand sides of
expression (\ref{ex_eq2}) for parameters $\kappa =0.52$ and
$r_*=8.0 \ h^{-1}$ Mpc are plotted. With these parameters we get

\begin{equation}
b(r)=\left(\frac{r}{8.0 \ h^{-1}\ {\rm Mpc}}\right)^{0.52},
\end{equation}
or, using a different expression

\begin{equation}
\xi _{\rho \rho}=b^{-2}(r)\xi (r)
=\left(\frac{r}{5.3 h^{-1}\ {\rm Mpc}}\right)^{-2.81}
\end{equation}
in the range above defined for $\xi$,
where $<\rho \rho>(r)=<n>^2(1+\xi _{\rho \rho} (r))$. The obtained power law
for mass correlation, ``the $-2.8$ power law'' is different from
that of objects' correlation, ``the $-1.8$ power law''. 
Of course, the left and hand sides of (\ref{ex_eq2}) were
not expected to agree perfectly and in fact they do not (Fig. 2) because
we assumed a $b$ dependence, a power law, which might not be very realistic. A
worthwhile result is that a deviation from $b=1$ is necessary to get an agreement
between the left- and right-hand sides of (\ref{ex_eq2}), and the solution must
be close to $b(r)=0.34r^{0.52}$. 
This result increases the information about the mass-mass
correlation function (see Peebles 1980), which was unknown till now. If this
result is true and is verified by alternative methods, it would give us
important information about the ``dynamics'' in the large scale distribution
of matter in the Universe.

The dependence of the outcome on the cut-off value is
not negligible, but neither is it very pronounced. Some numerical
results were obtained with other values of the cut-off, and the
qualitative result does not differ too much: a cut-off at
$r=0.1 \ h^{-1}$ Mpc instead of $r=0.3 \ h^{-1}$ Mpc, gives us
 $\kappa =0.47$ instead of $0.52$, and $r_*=8.1\ h^{-1}$ Mpc instead
of $r_*=8.0\ h^{-1}$ Mpc.

\begin{figure}
\begin{center}
\mbox{\epsfig{file=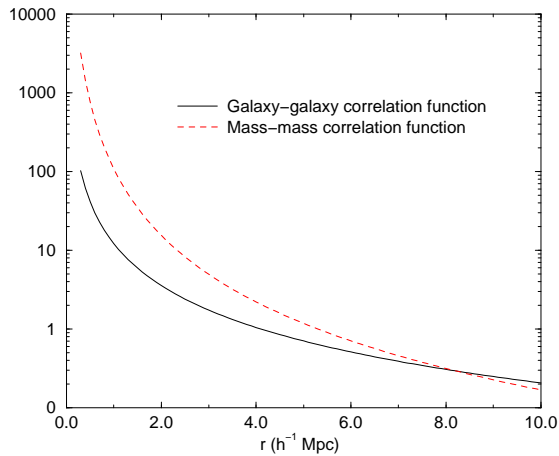,height=7cm}}
\end{center}
\caption{The galaxy-galaxy correlation function, $\xi $, derived
from observational data and $\xi _{\rho \rho}$, mass-mass correlation
function, derived from $\xi $ through
the force equation explained in this paper.}
\end{figure}

The direct 
consequence of this is that mass is more correlated than the objects
(Fig. 3).
This means that dark matter must exist near galaxies and clusters of galaxies to
increase the density contrast for short distances from an object
(dark matter in form of discrete unseen galaxies, for instance dwarfs galaxies),
unless the most massive galaxies are clumped together or there is some other solution, but something must explain the difference in both correlation
functions for galaxies and mass. Deviations from equilibrium
might also be responsible for part of the deviation attributed to
biasing, as far as it is only an approximation. I am unware of
how this may affect my results, essentially because the extent of this deviation
is unknown,
though I expect it to be not very large. A large
departure from equilibrium would produce a fast evolution of the distribution
and the distributions of matter far from us 
would have different distributions. This latter is not observed for long-range correlations
\footnote{Broadhurst et al. (1990) deduces from observational data a regular distribution
of galaxies similar to the nearby structure
up to distances of
$\sim 1000\ h^{-1}$ Mpc (a back in time further away from
$\sim 4\times 10^9$ yr)}, so
we must infer that the departure from equilibrium conditions cannot be too high
unless perhaps for very small ranges.
For typical scales in galaxy clusters, we also find strong evidence
of equilibrium (Carlberg et al. 1997). This discussion 
is beyond the scope of this paper. In Saslaw \& Hamilton (1984, their Sect. 6), we find further arguments in favour of this: ``Gravitational $N$-body experiments give good agreement with the theory.
This shows that even though an equilibrium theory may not explain the ultimate fate of galaxy clustering, it does provide a good description over extremely long time scales when the correlations are `frozen out' by the expansion of the Universe...[T]he equilibrium theory seems to explain most of the results.''

The main question is how equilibrium
could be reached in a short lifetime of the Universe. 
Violent relaxation which  enormously decreases the
relaxation time is a possible solution. Indeed, Saslaw (1985) points
out that this must be the mechanism that governs the system due to
large-scale collective modes
(see chapter 38 of Saslaw 1985) and Henriksen \& Widrow (1997) makes numerical simulations
achieving this. The consideration of a steady state in
the large-scale structure or a relaxation included in the initial 
conditions of the large-scale structure dynamics, before the formation of the
galaxies, are other possible explanations.

The deviation from equilibrium and the corrections to make to our equations
to take  these effects into account are topics for future papers. Further research is necessary in this area
to render these results more accurate.
With this example, we wanted to show the way of working with the
expressions described in this paper.

Readers might ask why I have not used the force equation to derive
the two point-correlation function. To do this, I would need a knowledge
of the biasing first, I cannot derive both things at the same time.
Since the two point correlation function is better known that
the bias, I decided to apply the method as described above. I had previously done some calculations to derive the two-point correlation
function assuming non-biasing and the result was not compatible
with observations, so I rejected this hypothesis thinking that
biasing is necessary as I obtained here.

\section{Results, other applications and further commentaries}

We have a numerical relationship between the distribution of galaxies
in space, which is represented by the correlation functions, and the
two-galaxy interaction, which is represented by the potential energy.

We have obtained the equality (\ref{testeq_exp}) that must be followed 
by the distributions
under Hypotheses 1 to 5. Also, 
we have equality (\ref{forceeq}) for the correlation 
functions and any interaction force under Hypotheses 1, 2 and 3
(we could also obtain an expresion like this including the expansion of space
by means of the method explained in the subsection dedicated to the
expansion).

Equation (\ref{testeq_exp}) relates the distribution correlation with the
mean velocity of the galaxies by means of $C$ with (\ref{vmean}), and
I believe it will be useful for obtaining a parameter from others
that are already known in the distribution: the mean velocity from
a complete knowledge of the correlation functions, or an unknown
parameter in the correlation function from the rest of the data. We could even obtain
more than one parameter: two or three (in my opinion, more than three are too
many) that follow the equality between the right- 
and  left-hand sides of equation (\ref{testeq_exp}). 

When equality between the two-sides of equation (\ref{testeq_exp})
is unattainable, this will indicate that our hypotheses are unsuitable. 
Probably, the most doubtful hypothesis is the first, i.e. that of
Boltzmann equilibrium, and it is possible to verify the relaxation 
using this equation.
\footnote{The verification should also be carried out with other methods
because there exists some possibility of a verification of the equality
due to other causes than relaxation, but the non-verification
of the equality implies directly the non-equilibrium when the other
hypotheses are right.}

In a sufficiently evolved system, Boltzmann equilibrium is achieved because
the particle-points are classical particles and the probability of a
state in such a case is proportional to the number of
different states for each particle that preserves the number of particles
and the total energy (the reader is referred to any book dealing
with the foundations of statistical
mechanics, e.g. Tolman 1938). It is also true that after a long time, 
the systems become virialized, and many systems are known to be in
these conditions, although not all of them. 

Otherwise, equation (\ref{forceeq}), {\it the force equation},
is a tool for looking for the kind of interaction in a system 
following Hypotheses 1, 2 and 3 (as purposed in Goldman et al. 1992). 
Once we have derived all the parameters of the
distribution we can fit a shape for $V$, a two-body interaction
potential that fits the equality, with or without expansion. 
We could even obtain other unknown
parameters ($C$ for example).

In order to demonstrate what might be the caveats in the implementation of
the force equation, I developed a real example in the previous section 
where it was used to infer information about the mass-mass correlation
function from the galaxy-galaxy correlation function, and the average
density and peculiar velocity in the large-scale distribution of
galaxies in the Universe. Further improvements are necessary, both in
the observations and the theoretical assumptions, to obtain an accurate
result, but the method is at least capable of telling us that the mass
is more correlated than the galaxies at short distances (Fig. 3) when
we assume relaxation on scales greater than $1\ h^{-1}$ Mpc.

\begin{acknowledgements}

I acknowledge gratefully the important and helpful 
comments made by J. Betancort-Rijo and the anonymous referee.
I also thank Monica Murphy and Terry Mahoney for a detailed
revision of the script.

\end{acknowledgements}

\end{document}